\documentclass[preprint,12pt]{elsarticle}

\usepackage[numbers]{natbib}
\usepackage{siunitx}
\usepackage{booktabs}
\usepackage{caption}
\usepackage{subcaption}
\usepackage{float}
\usepackage{tabularx}
\usepackage{adjustbox}
\usepackage{amsmath}
\usepackage{algorithm}
\usepackage{algpseudocode}
\usepackage{hyperref}
\usepackage{xcolor}
\usepackage{acronym}
\usepackage{multicol}
\usepackage{array}
\usepackage{makecell}
\usepackage{amssymb}

\journal{}

\begin{document}
\let\WriteBookmarks\relax
\def\floatpagepagefraction{1}
\def\textpagefraction{.001}

\renewcommand{\topfraction}{0.9}
\renewcommand{\bottomfraction}{0.8}
\renewcommand{\textfraction}{0.07}
\renewcommand{\floatpagefraction}{0.7}
\setcounter{topnumber}{3}
\setcounter{bottomnumber}{2}
\setcounter{totalnumber}{4}

\title{An AI-Driven Thermal-Fluid Testbed for Advanced Small Modular Reactors: Integration of Digital Twin and Large Language Models}

% Authors
\author[1]{Doyeong Lim\corref{cor1}}
\ead{dylim@tamu.edu}
\author[1]{Yang Liu\corref{cor1}}
\ead{y-liu@tamu.edu}
\author[1]{Zavier Ndum Ndum}
\author[1]{Christian Young}
\author[1]{Yassin Hassan}

% Affiliations
\affiliation[1]{
    organization={Department of Nuclear Engineering, Texas A\&M University},
    city={College Station},
    postcode={77843},
    state={TX},
    country={USA}
}

\cortext[cor1]{Corresponding author}

\begin{abstract}
This paper presents a multipurpose artificial intelligence (AI)-driven thermal-fluid testbed designed to advance Small Modular Reactor technologies by seamlessly integrating physical experimentation with advanced computational intelligence. The platform uniquely combines a versatile three-loop thermal-fluid facility with a high-fidelity digital twin and sophisticated AI frameworks for real-time prediction, control, and operational assistance. Methodologically, the testbed’s digital twin, built upon the System Analysis Module code, is coupled with a Gated Recurrent Unit (GRU) neural network. This machine learning model, trained on experimental data, enables faster-than-real-time simulation, providing predictive insights into the system’s dynamic behavior. The practical application of this AI integration is showcased through case studies. An AI-driven control framework where the GRU model accurately forecasts future system states and the corresponding control actions required to meet operational demands. Furthermore, an intelligent assistant, powered by a large language model, translates complex sensor data and simulation outputs into natural language, offering operators actionable analysis and safety recommendations. Comprehensive validation against experimental transients confirms the platform's high fidelity, with the GRU model achieving a temperature prediction root mean square error of 1.42 K. This work establishes an integrated research environment at the intersection of AI and thermal-fluid science, showcasing how AI-driven methodologies in modeling, control, and operator support can accelerate the innovation and deployment of next-generation nuclear systems.
\end{abstract}

% Keyword

\maketitle

\section{Introduction}
\label{sec:intro}

The evolution of nuclear energy systems toward greater flexibility and intelligence demands comprehensive research platforms that can bridge the gap between theoretical advances and practical implementation \citep{iea2022world}. As the energy landscape undergoes rapid transformation with increasing renewable penetration and evolving grid requirements, the nuclear industry faces pressing needs for experimental validation of advanced concepts ranging from innovative thermal-fluid designs to flexible operation capability and advanced strategy integration \citep{zhang2024load}. Traditional nuclear research facilities, often designed for single-purpose investigations, lack the versatility and cyber-physical integration necessary to address these multifaceted challenges. This limitation has created a critical gap in the development pipeline for advanced small modular reactors (SMRs) and their associated technologies \citep{chang2022review, wang2024small}.

The emergence of cyber-physical systems represents a fundamental paradigm shift in engineering research and development, offering opportunities to accelerate innovation through the seamless integration of physical experimentation, computational modeling, and Artificial Intelligence (AI) and Machine Learning (ML)~\citep{baheti2011cyber, liu2017review}. In the nuclear domain, this integration manifests as sophisticated testbeds that combine representative thermal-fluid facilities with digital twins, machine learning algorithms, and advanced human-machine interfaces. Such cyber-physical platforms enable researchers to investigate complex phenomena, validate computational models, develop advanced control strategies, and explore novel operational concepts within a unified experimental framework \citep{ayodeji2023cyber}. 

The nuclear industry has achieved breakthrough applications of digital twin technology, marking a transformative shift from traditional simulation approaches. Stewart et al. \citep{stewart2025agn} successfully demonstrated the research reactor digital twin on the AGN-201, achieving real-time data integration with predictive maintenance capabilities. Liu et al. \citep{liu2024development, liu2022sam} advanced the field further by developing Graph Neural Network and System Analysis Module (SAM)-based digital twins. This framework treats reactors as heterogeneous graphs, enabling faster-than-real-time simulations with accurate predictions based on limited sensor data. Lim et al. \citep{lim2025hybrid} further contributed with hybrid surrogate modeling approaches. Other recent digital twin application works include comprehensive architectural frameworks and remote monitoring systems \citep{mengyan2024current, stevens2023digital, ndum2025LFRDT}. However, a critical research gap persists between these digital twin implementations and their integration with advanced AI decision support systems for physical testbeds. While recent efforts have begun addressing this gap through platforms, most existing digital twins remain isolated from comprehensive physical validation and lack sophisticated AI-driven autonomous control capabilities \citep{barik2024achieving}.

Recent progress has occurred on two parallel but largely disconnected fronts. Experimental side, the MARVEL microreactor at Idaho National Laboratory represents the first new reactor at INL in over 40 years and is specifically designed for autonomous control testing and microgrid integration research \citep{patterson2024marvel}. Similarly, Kairos Power demonstration reactor marks the first Generation IV reactor approved for construction by the NRC, utilizing innovative TRISO fuel and molten fluoride salt cooling technologies \citep{rykhlevskii2024kairos, satvat2025hermes}. Prantikos et al. \citep{prantikos2023physics} at Purdue University achieved another milestone as the first fully digital nuclear reactor in the United States, featuring complete digital instrumentation and control systems that enable comprehensive testing of AI-powered simulation and cybersecurity frameworks. Despite these advances, existing testbeds primarily focus on single-technology demonstrations rather than integrated cyber-physical systems.

Concurrently, the digital domain has seen a surge of AI and machine learning (ML) applications that significantly enhance modeling and simulation (M\&S) for nuclear thermal-fluids. Researchers have successfully developed data-driven surrogate models and physics-informed neural networks to predict complex phenomena \citep{pirnstill2025surrogates}, from flow dynamics \citep{weng2025property} to advanced boiling heat transfer\citep{dunlap2025hit2flux, lim2025acoustic}, often serving as advanced closure models for system-level codes like SAM \citep{lanade2025high, liu2018data}. Another critical area of advancement is uncertainty quantification (UQ), where Bayesian frameworks and deep learning are used to assess the confidence in simulation results—a vital step for making informed, safety-conscious decisions in nuclear engineering \citep{liu2023uncertainty, abulawi2025bayesian}. 

Despite these significant advances in both physical hardware and sophisticated digital modeling, a critical research gap persists at their intersection. Most experimental facilities remain focused on single-technology demonstrations, while the powerful AI-driven M\&S and UQ techniques are often developed in isolation, lacking comprehensive physical validation and integration into sophisticated, real-time control and decision-support architectures for physical systems \citep{barik2024achieving}. The key challenge, therefore, is the seamless fusion of these advanced physical and digital components into a single, intelligent research platform.

To address this integration gap, this study presents the development and comprehensive validation of an AI-driven thermal-fluid testbed designed as a versatile research platform for advancing SMR technologies and their applications. The testbed uniquely integrates three essential components: (1) A flexible three-loop thermal-fluid facility as physical testbed with comprehensive instrumentation enabling investigation of diverse phenomena and operational scenarios; (2) A digital twin based on SAM (System Analysis Module) code \citep{hu2021sam} - advanced nuclear thermal-fluid system analysis code - providing simulation and virtual sensing capabilities; (3) Advanced AI integration including prediction of future state and control actuation, and large language models for intelligent system interaction. This multipurpose platform supports a wide spectrum of research activities, from fundamental thermal-fluid studies and model validation to autonomous control development and human factors engineering. Through multiple case studies, we aim to demonstrate how this thermal-fluid testbed addresses critical research needs across different domains. 

The remainder of this paper is organized as follows: Section~\ref{sec:testbed} presents the comprehensive design and implementation of the AI-driven thermal-fluid testbed, detailing the thermal-fluid facility design, digital twin integration methodology, and AI capabilities. Section~\ref{sec:case_studies} demonstrates the platform's versatility through diverse case studies, including computational model validation, prediction of state and control actuation implementation, LLM implementation, and other potential applications. Section~\ref{sec:conclusions} discusses the broader implications for nuclear technology development and outlines future expansion of this multipurpose research platform.

%%%%%%%%%%%%%%%%%%%%%%%%%%%%%%%%%%%%%%%%%%%%%%%%%%%%%%%%%%%%%%%%%%%%%%%%%%%%%%%%%%%%
\section{AI-Driven Thermal-Fluid Testbed}
\label{sec:testbed}

This section details the architecture of the multipurpose AI-driven testbed, a cyber-physical platform designed for advanced nuclear research. The system's architecture is built on three deeply integrated components, each described in the following subsections: (1) a three-loop thermal-fluid facility with comprehensive instrumentation and control that serves as the physical asset, (2) a high-fidelity digital twin that provides predictive simulation, and (3) an AI-driven layer, including a Large Language Model, that enables intelligent control and operational assistance.

\subsection{Thermal-Fluid Facility}
\label{subsec:facility}

\subsubsection{Design characteristics}

The experimental thermal-fluid facility was developed as a versatile platform to investigate diverse phenomena relevant to advanced small modular reactors while maintaining operational flexibility for various research applications. Drawing inspiration from molten salt reactor configurations, particularly the Kairos Power fluoride salt-cooled high-temperature reactor (KP-FHR)~\citep{blandford2020kairos}, the testbed incorporates top-mounted pumps and operates under near-atmospheric pressure conditions. While referenced MSR designs typically employ pebble-bed fuel configurations, our experimental setup utilizes a streamlined rod configuration featuring four cartridge-type electrical heaters positioned within the primary loop's transparent test section. The facility design prioritizes modularity, scalability, and thermal-fluid behavior to enable comprehensive investigations in multiple research domains.

\begin{figure}[htbp]
    \centering
    \includegraphics[width=\textwidth]{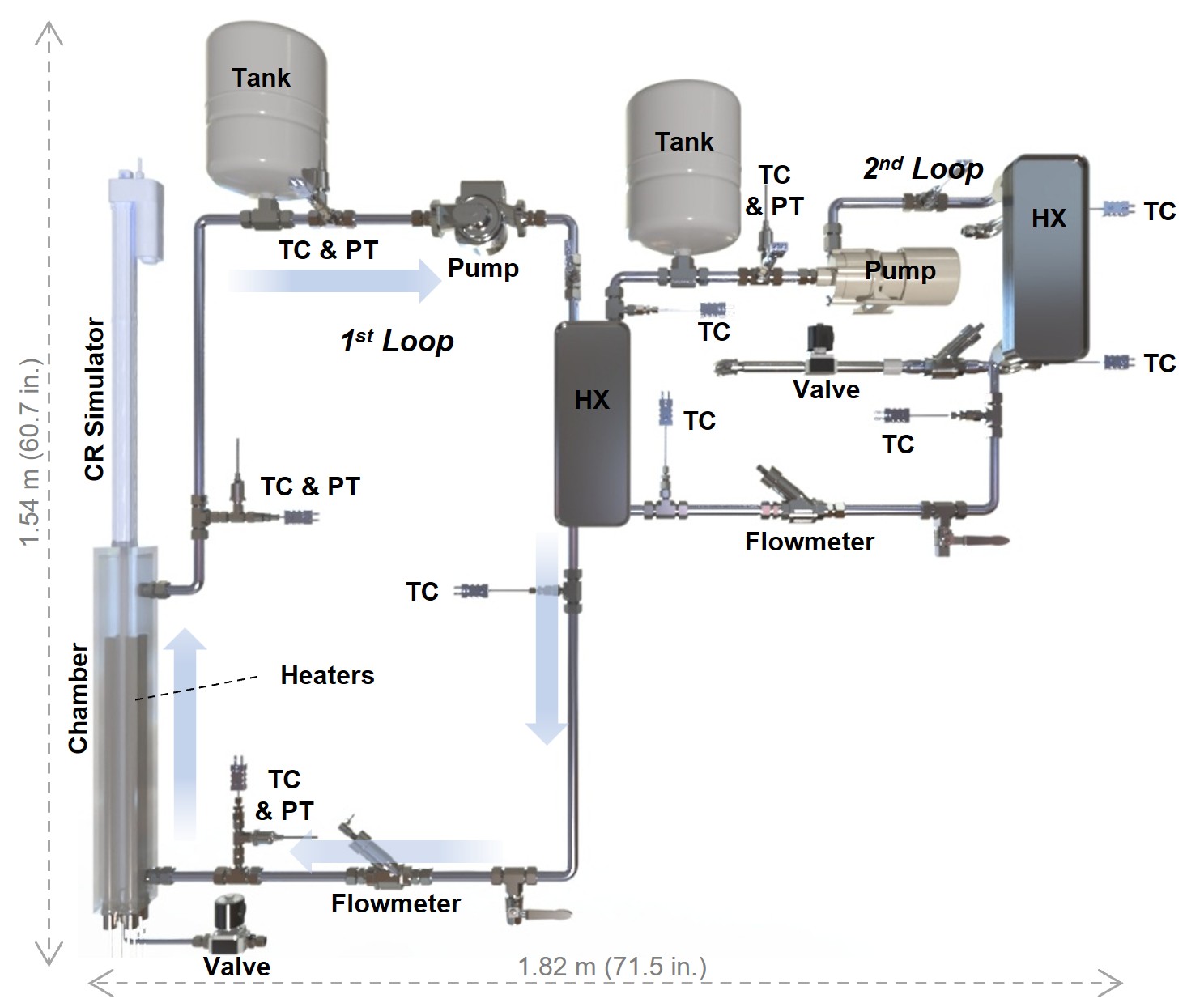}
    \caption{Schematic of the thermal-fluid test facility as a physical testbed, illustrating the primary, secondary, and tertiary loops and key components.}
    \label{fig:facility_design}
\end{figure}

Figure~\ref{fig:facility_design} presents the integrated testbed architecture through complementary perspectives for the detailed CAD representation, revealing spatial arrangements and component interfaces. The facility comprises three independent thermal-fluid loops interconnected through brazed plate heat exchangers, enabling compact design with high heat transfer rates and controlled temperature distributions across the system.

The primary loop serves as the heat generation section, incorporating four controlled cartridge heaters installed within a transparent polycarbonate chamber (the maximum fluid temperature to \SI{120}{\celsius}). This transparent design enables direct visual observation of flow dynamics and potential flow instabilities during various operational conditions. The heaters are arranged in a 2$\times$2 square lattice configuration, providing uniform heat flux distribution while maintaining accessibility for various experimental configurations. Each cartridge heater features a maximum power rating of \SI{3.925}{\kilo\watt} with \SI{208}{\volt}AC, constructed with Inconel sheathing to ensure thermal stability and corrosion resistance. The heaters are controlled through individual solid-state relays interfaced with a master power controller (DIN-A-MITE, Watlow), enabling independent power regulation. The total heating capacity of \SI{15.7}{\kilo\watt} provides sufficient thermal driving force to achieve representative temperature rises.

A motorized control rod simulator (PA-14P, Progressive) was placed on the top of the transparent polycarbonate chamber and centered on the four cartridge heaters. This control rod simulator features a precision linear actuator capable of \SI{50.8}{\milli\meter\per\second} traverse speeds across the full \SI{609.6}{\milli\meter} stroke length, which was enough for full insertion onto bottom of the heater. The control rod mechanism employs a stepper motor, monitored through an integrated potentiometer providing real-time position feedback. This assembly enables realistic reactivity control studies and supports investigations of control rod worth, insertion dynamics, and automated reactivity management strategies.

The secondary and tertiary loops provide progressive heat removal stages, with a heat exchanger (GBS 400H, Kelvion) between loops occurring through brazed plate heat exchangers selected for their compact design, high thermal effectiveness, and minimal pressure drop characteristics. Both primary and secondary loops feature variable-speed centrifugal pumps capable of \SI{0}{\sim}\SI{0.76}{\kilo\gram\per\second} flow rates. The pumps (TACO Model 0011 for primary, Little Giant 5-MD for secondary) are strategically positioned at the top of each loop to facilitate natural circulation initiation, accommodate potential gas entrainment. Each pump assembly includes variable frequency drives (VFDs, ATO GK3000) with \SI{0.1}{\hertz} resolution, enabling precise flow control and supporting studies of natural circulation transitions.

All piping systems employ \SI{19}{\milli\meter} (3/4-inch) stainless steel tubing with comprehensive thermal insulation to minimize heat losses and ensure accurate energy balance measurements. The insulation method utilizes fiberglass pipe wrap, maintaining external surface temperatures close to ambient conditions. The facility accommodates both water and surrogate heat transfer fluids (e.g., DOWTHERM SR-1) for molten salt fluid under near-atmospheric pressure conditions, enabling investigations across temperature ranges relevant to advanced molten salt reactors and light water SMRs.

\subsubsection{Instrumentation and Control}

The comprehensive instrumentation enables real-time monitoring, data acquisition, and control across all facility subsystems. Table \ref{tab:parameters} summarizes the measurement and control specifications for key sensors and controllers of the testbed. Temperature measurements utilize strategically positioned thermocouples providing comprehensive thermal mapping: 12 T-type thermocouples ($\pm$\SI{1.0}{\celsius} uncertainty) positioned at loop inlets, outlets, and intermediate locations, and 4 J-type thermocouples ($\pm$\SI{2.2}{\celsius} uncertainty) embedded within cartridge heater assemblies for overheat protection and performance monitoring.

Flow measurements employ variable area flow transmitters with $\pm$\SI{5.0}{\percent} accuracy across the operational range. Each loop incorporates dedicated flow sensors positioned in the inlet of the chamber in the primary loop, the inlet of heat exchangers in the secondary and the tertiary loop, with straight pipe sections to ensure measurement accuracy. The sensors feature 4--20~mA analog outputs with digital communication capability. Pressure measurements utilize gauge pressure transducers ($\pm$\SI{1.0}{\percent} accuracy) positioned at the inlet and outlet of chamber, top pipe in the primary loop, and top pipe position in the secondary loop.

\begin{table}[htbp]
    \caption{Measurement and control specifications of the testbed.}
    \label{tab:parameters}
    \centering
    \small
    \begin{tabular}{@{}llll@{}}
        \toprule
        \textbf{Sensor/Controller} & \textbf{Parameter} & \textbf{Max. Value} & \textbf{Uncertainty} \\
        \midrule
        T-type TC & Fluid Temperatures & \SI{220}{\celsius} & $\pm$\SI{1.0}{\celsius} \\         
        J-type TC & Heater Internal Temp. & \SI{750}{\celsius} & $\pm$\SI{2.2}{\celsius} \\
        Flow Meter & Loop Flow Rates & \SI{0.76}{\kilo\gram\per\second} & $\pm$\SI{5.0}{\percent} \\
        Pressure Transducer & Gauge Pressure & \SI{206.8}{\kilo\pascal} & $\pm$\SI{1.0}{\percent} \\
        Voltage Sensor & Heater Voltage & \SI{250}{\volt}AC & $\pm$\SI{0.5}{\percent} \\
        Current Sensor & Heater Current & \SI{100}{\ampere} & $\pm$\SI{0.5}{\percent} \\
        Heater Controller & Applied Power Control & \SI{15.7}{\kilo\watt} & $\pm$\SI{5.0}{\percent} \\
        Pump Controller & Pump Speed Control & \SI{60}{\hertz} & $\pm$\SI{1.0}{\percent} \\
        CR Controller & Control Rod Position & \SI{609.6}{\milli\meter} & $\pm$\SI{5.0}{\percent} \\
        \bottomrule
    \end{tabular}
\end{table}

The testbed employs a distributed control architecture that integrates industrial-grade hardware through a PLC system (R.Stahl I.O Module), providing \SI{10}{\hertz} sampling rates with deterministic control response times below \SI{100}{\milli\second}. This system provides three primary control functions: (1) regulation of heater power via AC voltage control with resolution of \SI{5.0}{\percent} throughout the whole range \SI{15.7}{\kilo\watt}, (2) control of pump flow through variable frequency drives that operate up to \SI{60}{\hertz}, and (3) positioning of the control rod using closed-loop stepper motor control with potentiometer feedback achieving $\pm$\SI{5.0}{\percent} positioning accuracy.

\subsubsection{Communication Infrastructure}

\begin{figure}[htbp]
    \centering
    \includegraphics[width=\textwidth]{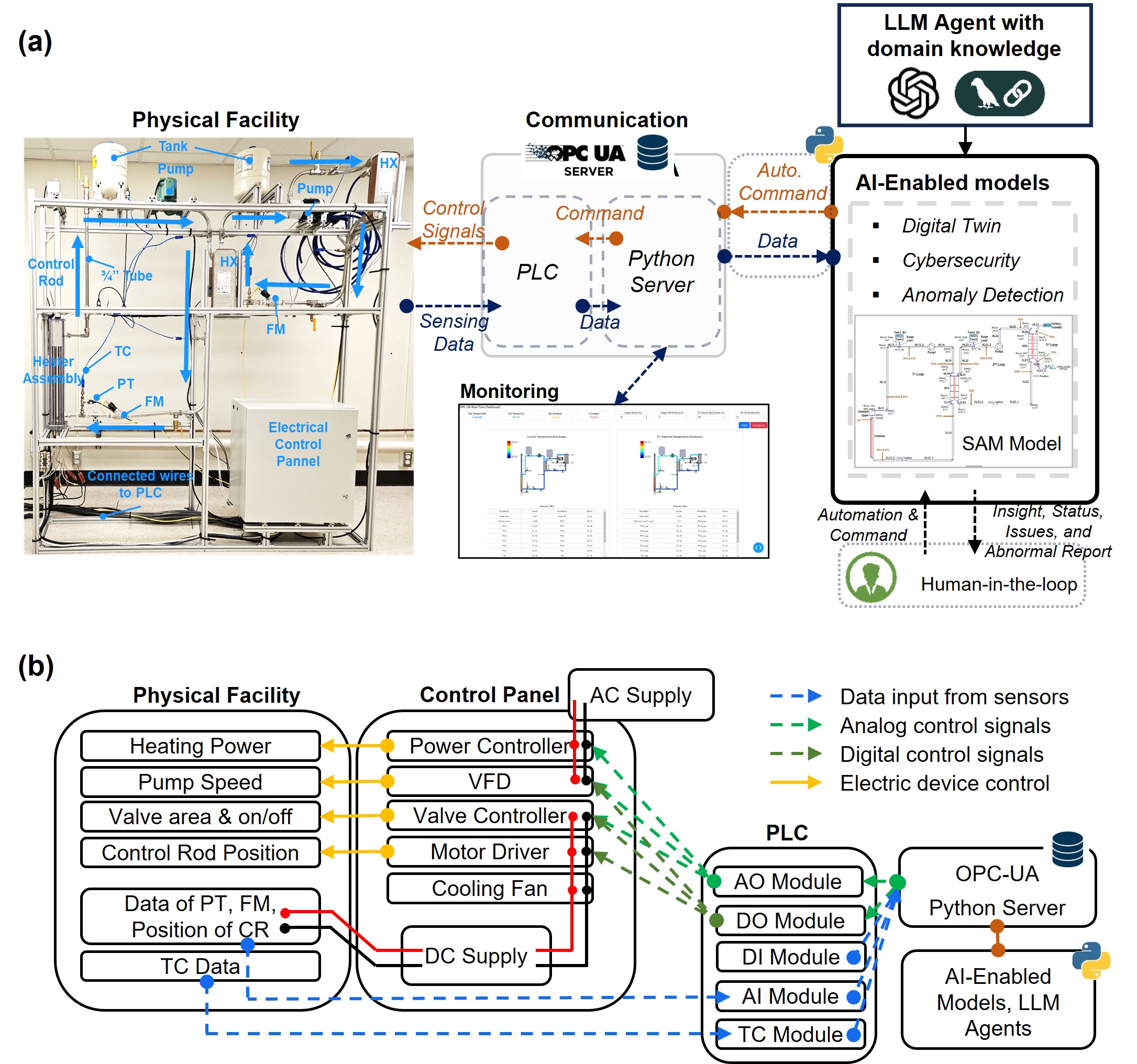}
    \caption{(a) Integrated thermal-fluid architecture showing the physical facility, OPC-UA communication infrastructure, SAM-based digital twin, and AI framework integration, and (b) schematic communication and control pathway within the testbed.}
    \label{fig:architecture}
\end{figure}

The thermal-fluid integration of the testbed relies on a robust communication infrastructure that bridges the physical facility with computational models and AI systems. Figure~\ref{fig:architecture} illustrates the complete system topology, encompassing sensor networks, communication protocols, and computational components that enable real-time synchronization between physical and virtual domains.

The communication backbone employs OPC-UA (Open Platform Communications Unified Architecture) protocol, selected for its deterministic performance, built-in security features, and widespread adoption in industrial automation. The OPC-UA server, implemented using the open62541 library, establishes a unified namespace encompassing all sensor readings, actuator states, and control variables. This server maintains a sampling rate of \SI{10}{\hertz} for critical parameters (temperatures, flow rates, heater power) and \SI{1}{\hertz} for auxiliary measurements, ensuring adequate temporal resolution while minimizing network overhead.

External system integration utilizes OPC-UA's service-oriented architecture, enabling seamless connection with Python-based control algorithms, machine learning models, and digital twin implementations. The Python control server, developed using the asyncua library, serves as the central intelligence hub, maintaining the \SI{100}{\milli\second} control loop timing required for stable operation while enabling integration of advanced algorithms. The PLC retains essential safety interlocks and low-level device control, with complex decision-making delegated to higher-level systems.

%%%%%%%%%%%%%%%%%%%%%%%%%%%%%%%%%%%%%%%%%%%%%%%%%%%%%%
\subsection{Integration with Digital Twin}
\label{subsec:digital_twin}

\begin{figure}[htbp]
\centering
\includegraphics[width=\textwidth]{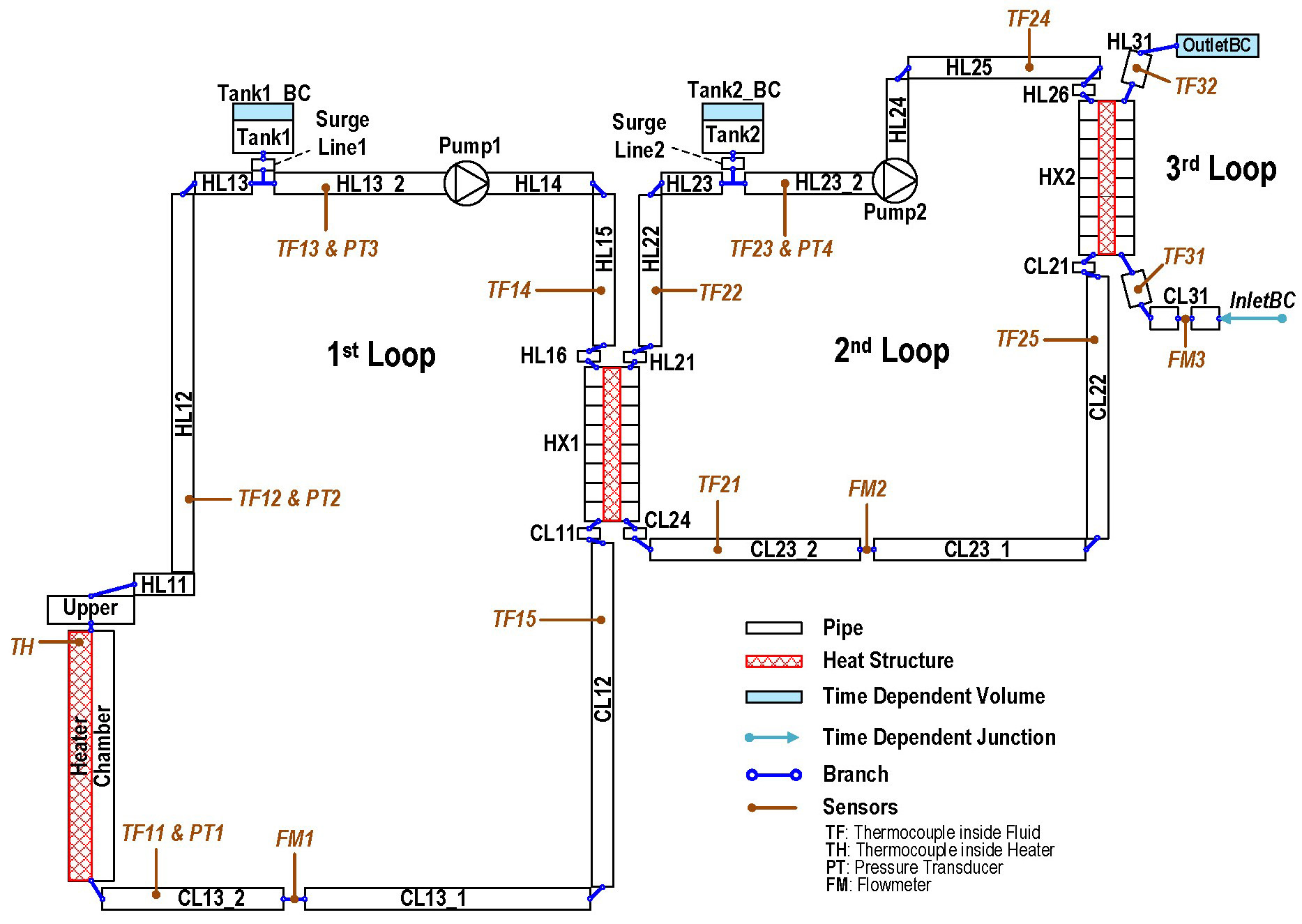}
\caption{SAM nodalization model for digital twin implementation of the thermal-fluid testbed.}
\label{fig:sam_nodal}
\end{figure}

The digital twin leverages the System Analysis Module (SAM) code to provide robust thermal-fluid simulation of the three-loop testbed. As shown in Figure~\ref{fig:sam_nodal}, the SAM model implements a comprehensive nodalization scheme with 127 fluid nodes and 5 heat structures with water property provided by the MOOSE framework, enabling accurate capture of transient phenomena and spatial temperature distributions.

The heating chamber in the primary loop is modeled using a PBCoreChannel component representing four cartridge heaters as cylindrical heat structures with 0.455 m active length divided into 15 axial elements. Key geometric parameters include a hydraulic diameter of 16.8 mm, heat transfer surface density of 127.8 m²/m³, and pitch-to-diameter ratio of 1.43. All piping networks utilize PBPipe components with integrated wall conduction modeling.

Heat exchangers employ counter-flow PBHeatExchanger components with compact geometry specifications: 2.0e-4 m² flow areas, 5 mm hydraulic diameter, and enhanced heat transfer surface density. Flow control components include variable-speed pumps calibrated against manufacturer data (TACO Model 0011 for primary, Little Giant 5-MD for secondary) and flow meters represented as variable K-factor branches dependent on Reynolds number.

The SAM model calibration involved systematic parameter optimization against experimental data across multiple operational scenarios. The calibration process employed a multi-objective optimization approach targeting temperature predictions, flow distribution accuracy, and pressure drop matching. Key calibration parameters included heat transfer coefficients, friction factors, heat loss coefficients, and pump characteristics.

The validation methodology comprised steady-state and transient testing campaigns. Steady-state validation involved comparing predicted temperatures, flow rates, and power against experimental measurements across the operational envelope. Transient validation evaluated the model's ability to predict dynamic response during power ramp scenarios. The integration of AI acceleration techniques with this validated SAM model enables real-time digital twin capabilities while maintaining high fidelity simulation accuracy.

In this study, the SAM model serves as the foundation for our digital twin implementation. Given that digital twins require real-time information provision capabilities, we have integrated AI-based computational acceleration techniques to enhance the model's responsiveness for experimental facility monitoring and digital twin applications.

\subsection{Integration with Large Language Model}
\label{subsec:llm}

The integration of a Large Language Model (LLM) into the testbed represents a meaningful advancement in real-time monitoring, diagnostic capabilities, and operational decision-making for nuclear facilities. The LLM functions as an intelligent interface between complex sensor data and human operators, transforming multi-dimensional operational parameters into actionable insights through natural language processing.

\begin{figure}[htbp]
    \centering
    \includegraphics[width=\textwidth]{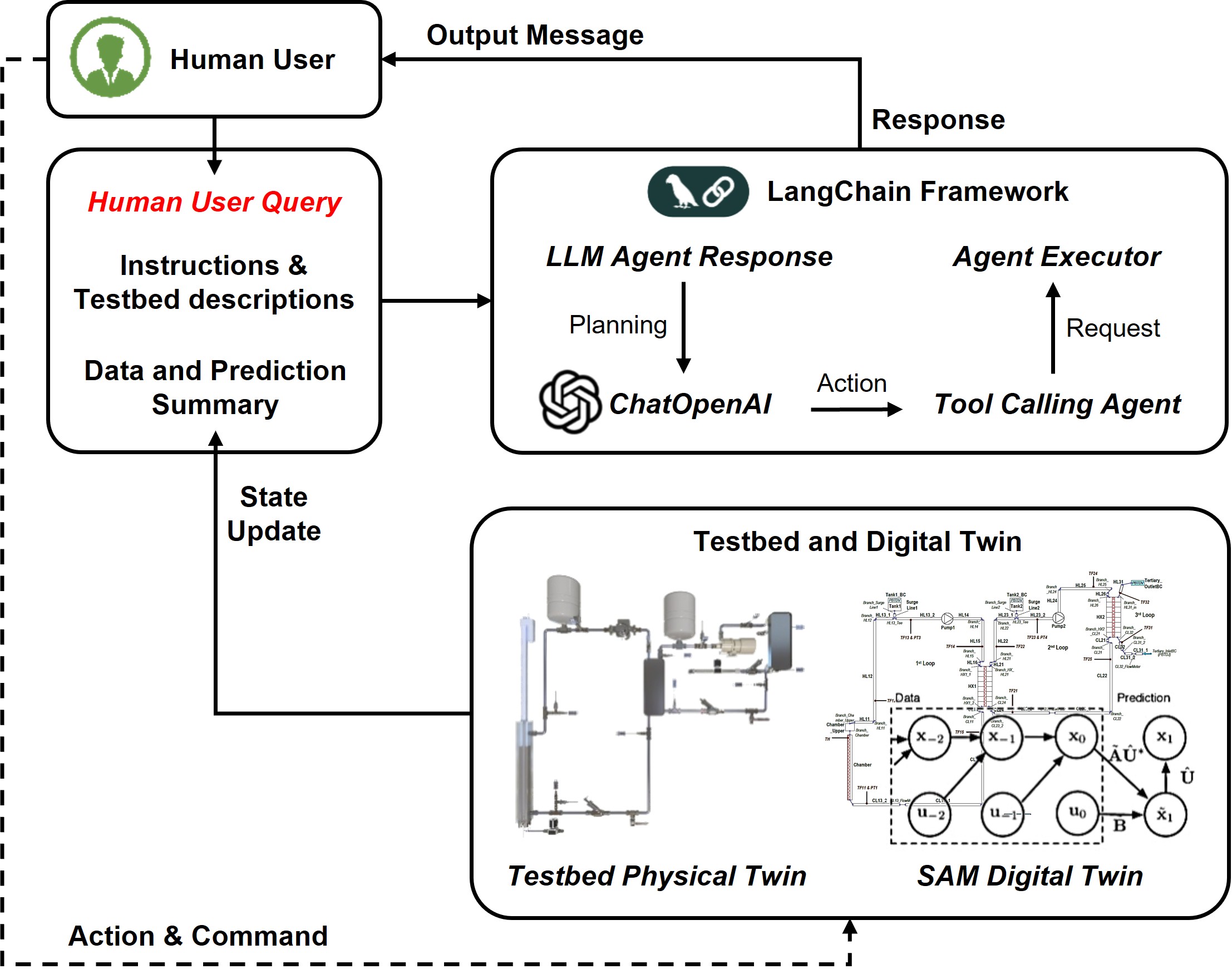}
    \caption{An overview of the LLM-based workflow for a thermal-fluid testbed and its digital twin using ChatGPT (OpenAI) and the LangChain framework.}
    \label{fig:llm_workflow}
\end{figure}

Figure~\ref{fig:llm_workflow} illustrates the comprehensive workflow for LLM-based operational assistance in the thermal-fluid testbed and its digital twin. The system architecture comprises four main components interconnected through a cyclical data flow pattern. At the top level, the Human User interacts with the system through natural language queries, initiating the analytical process. These queries, along with testbed descriptions and operational instructions, form the Human User Query component, which continuously receives state updates from the physical system to maintain current operational awareness. The query processing flows to the LangChain framework, which serves as the management layer for AI-driven analysis. Within this framework, the LLM Agent Response module, powered by ChatGPT (OpenAI), performs two functions: Planning, where it interprets the user's intent and determines the appropriate analytical approach, and Action, where it executes the analysis through the Tool Calling Agent. The Agent Executor manages these interactions, coordinating between the LLM and various analytical tools as needed. The bottom layer represents the Testbed and Digital Twin infrastructure, consisting of the physical thermal-fluid testbed and its corresponding SAM Digital Twin. The workflow operates as a closed-loop system where user queries trigger LLM analysis, which can result in actionable commands being sent back to the testbed (shown as "Action \& Command" in the dashed feedback line). The system simultaneously provides human-readable responses as "Output Messages," completing the interaction cycle and enabling informed operational decisions.

\begin{table}[htbp]
    \centering
    \caption{Large language model integration algorithm for operational and monitoring assistance of experimental testbed.}
    \label{tab:llm_algorithm}
    \small
    \begin{tabularx}{\textwidth}{|>{\hsize=0.5\hsize\centering\arraybackslash}X|>{\hsize=1.5\hsize}X|}
        \hline
        \textbf{Step} & \textbf{Description} \\[0pt]
        \hline
        \textbf{1. Query Acquisition} & 
        \texttt{Input: Natural language query from operator interface\newline query ← getUserInput(dashboard\_interface)}\\
        \hline
        
        \textbf{2. Data Acquisition} & 
        \texttt{Process: Synchronous collection from OPC-UA nodes\newline
        data\_dict ← readOPCUA(client, nodes[])}\\
        \hline
        
        \textbf{3. Parameter Computation} & 
        \texttt{Process: Calculate specific parameters\newline$P_{kW} = V \times I / 1000$ (Total power)\newline$P_{elec} = P_{kW} \times 0.45 $ (Electricity power)}\\
        \hline
        
        \textbf{4. Context Construction} & 
        \texttt{Process: Structure facility state for LLM\newline context ← formatFacilityState(data\_dict)\newline Constructs comprehensive state: electrical parameters, thermal vectors, hydraulic data, timestamp}\\
        \hline
        
        \textbf{5. Query Augmentation} & 
        \texttt{Process: Combine query with facility context\newline Ensures LLM has complete situational awareness}\\
        \hline
        
        \textbf{6. LLM Inference} & 
        \texttt{Process: Generate context-aware response \newline response ← llm\_agent\_response("gpt-4o", $Q_{aug}$)}\\
        & Provides: operational insights, anomaly detection, maintenance recommendations, safety analysis\\
        \hline
        
        \textbf{7. Response Delivery} & 
        \texttt{Output: Display analysis on dashboard \newline dashboard.display(response) \newline Real-time synchronized presentation of insights}\\
        \hline
    \end{tabularx}
\end{table}

The implementation of this workflow follows a systematic algorithm detailed in Table~\ref{tab:llm_algorithm}, which outlines the step-by-step process from query acquisition to response delivery. The algorithm begins with continuous data acquisition from OPC-UA nodes distributed throughout the facility's three thermal-fluid loops. These sensors monitor critical parameters, including temperatures across primary, secondary, and tertiary loops, heater surface temperatures, gauge pressures, volumetric flow rates, electrical parameters (voltage, current), and control rod position.

The system performs real-time calculations to derive essential operational metrics, as shown in Step 3, including total power consumption, electrical power efficiency, and average heater temperatures. These raw measurements and computed parameters are then structured into a comprehensive facility state representation through the context construction phase (Step 4). This structured data forms the foundation for the LLM's situational awareness, enabling it to understand the facility's operational state holistically rather than as isolated data points. When operators submit natural language queries through the web-based dashboard interface—such as "Analyze the current thermal distribution and advise if it is safe to increase reactor power"—the system augments these queries with the real-time facility context (Step 5), creating an information-rich prompt that combines human intent with current operational data. The LLM inference engine, powered by OpenAI's GPT-4o model~\citep{OpenAI2024GPT4o}, processes these augmented queries to generate context-aware responses across multiple operational domains.

%%%%%%%%%%%%%%%%%%%%%%%%%%%%%%%%%%%%%%%%%%%%%%%%%
\section{Case Studies}
\label{sec:case_studies}

The multipurpose nature of the AI-driven thermal-fluid testbed enables diverse research applications spanning thermal-fluid investigations to advanced development. This section presents multiple case studies demonstrating the versatility and research capabilities of the testbed.

\subsection{SAM Validation for Digital Twin}
\label{subsec:sam_validation}

The foundational element of the thermal-fluid testbed is the SAM-based digital twin, which must accurately replicate the thermal-fluid behavior of the physical facility. To validate its fidelity, experimental campaigns were conducted covering steady-state conditions and transient scenarios.

\begin{figure}[htbp]
\centering
\includegraphics[width=\textwidth]{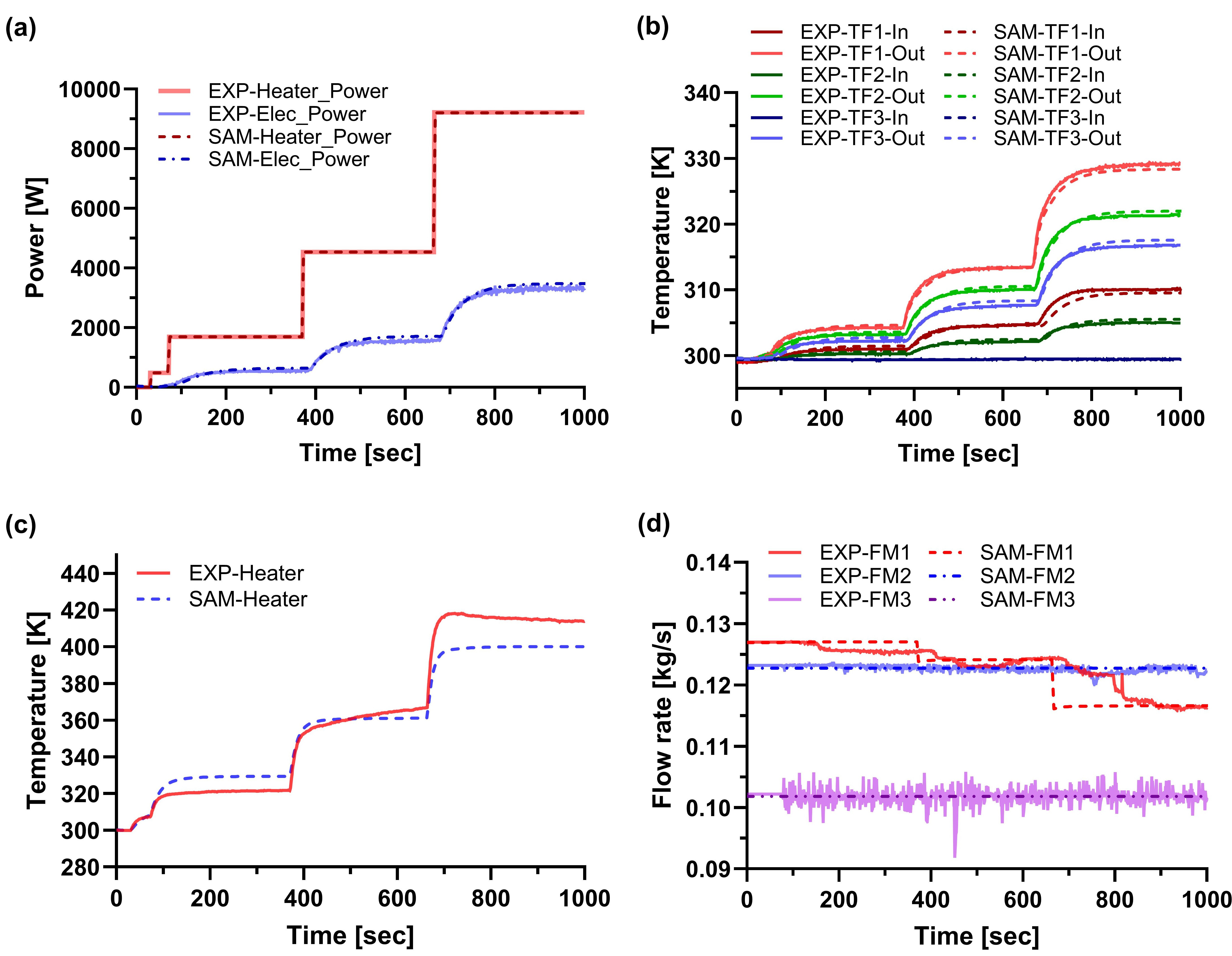}
\caption{SAM digital twin validation against experimental data showing: (a) applied heater power and estimated electrical power, (b) temperature responses at inlets and outlets of each loops, (c) heater temperature comparison, and (d) flow rate measurements during multi-step power transient from 0 to 9.2 kW.}
\label{fig:sam_validation}
\end{figure}

The validation campaign encompassed a multi-step power transient test designed to evaluate the digital twin's predictive capabilities across the facility's operational envelope. As illustrated in Figure~\ref{fig:sam_validation}, the experimental protocol initiated from ambient conditions and progressed through four distinct power levels: 490 W, 1692 W, 4536 W, and 9220 W, with each step maintained for a sufficient duration to achieve thermal equilibrium. This stepwise approach enabled a comprehensive assessment of both steady-state accuracy and transient response characteristics.

Model calibration yielded excellent agreement with experimental data: mean temperature prediction errors of 1.2°C, flow rate deviations under 2.8\%, and pressure drop predictions within ±4.2\% across all test conditions. The calibrated digital twin runs in real-time at 1 Hz, providing comprehensive state estimates that include 16 temperature measurements, 3 flow rates, and 4 pressure readings to support training of the neural network controller and autonomous control operations in real time.

The temperature prediction accuracy demonstrated exceptional fidelity throughout the test campaign. Analysis of the primary loop outlet temperature revealed a mean absolute error (MAE) of 0.667 K across all operational conditions, while the secondary loop outlet achieved an even lower MAE of 0.228 K. These results significantly exceed the initial validation target of 1.2 K accuracy. The digital twin's performance remained consistent across different power levels, with maximum instantaneous errors of 0.384 K observed during the initial power step from 0 to 490 W, attributed to the model's conservative heat loss estimations at low power conditions.

Transient response characterization revealed the digital twin's capability to accurately capture system dynamics during power transitions. The SAM model successfully predicted the exponential temperature rise characteristics, with time constants matching experimental observations within 8\% across all power steps. From 4536 W to 9220 W, representing a 103\% power increase, the model maintained temperature prediction errors below 0.4 K while accurately capturing the 280-second rise time to reach 95\% of the steady-state value. The slight temperature under prediction at maximum power (304.69 K simulated versus 305.07 K experimental) suggests conservative modeling of heat transfer enhancement at high Reynolds numbers, providing an inherent safety margin for control applications.

Flow rate predictions exhibited remarkable accuracy with an average error of 1.12\% across the entire test duration. The digital twin precisely captured the pump characteristic curve behavior, maintaining flow prediction errors below 2.8\% even during rapid thermal transients where density variations induced natural circulation effects superimposed on forced flow. The model's ability to predict flow redistribution between parallel channels during asymmetric heating scenarios further validated the momentum equation implementation and junction loss correlations.

The validated SAM digital twin provides comprehensive state estimates for 247 unmeasured parameters, including wall temperatures, fluid properties in each fluid element, and heat transfer coefficients. This virtual sensor capability enables advanced control strategies that would be infeasible with physical instrumentation alone. The model's demonstrated accuracy its robust performance during severe transients establish it as a reliable foundation for the thermal-fluid testbed's model-predictive control and real-time digital twin applications.

%%%%%%%%%%%%%%%%%%%%%%%%%%%%%%%%%%%%%%%%%%

\subsection{AI-driven Condition Prediction and Operation Control}
\label{subsec:ml_control}

This case study demonstrates the implementation of AI for real-time prediction of future conditions and control actions, and digital twin acceleration for real-time digital twin operation using Gated Recurrent Unit (GRU) neural networks trained on experimental and SAM simulation data.

\subsubsection{AI architecture and training method}

Recurrent neural networks (RNNs) are well suited to model time-dependent processes, yet classic RNN cells suffer from the vanishing-gradient problem when learning long-term dependencies. The Gated Recurrent Unit (GRU)~\citep{chung2014empirical} mitigates this issue by adding two learnable gates that regulate information flow while keeping the network lightweight enough for real-time deployment.

Let $\mathbf{x}_t\in\mathbb{R}^{d_x}$ denote the input and $\mathbf{h}_{t-1}\in\mathbb{R}^{d_h}$ the hidden state at time~$t-1$. With concatenation operator $[\cdot;\cdot]$ and element-wise product $\odot$, a GRU layer performs

\begin{align}
    \mathbf{r}_t &= \sigma\!\bigl(\mathbf{W}_r[\mathbf{h}_{t-1};\mathbf{x}_t]+\mathbf{b}_r\bigr),\\
    \mathbf{z}_t &= \sigma\!\bigl(\mathbf{W}_z[\mathbf{h}_{t-1};\mathbf{x}_t]+\mathbf{b}_z\bigr),\\
    \tilde{\mathbf{h}}_t &= \tanh\!\bigl(\mathbf{W}_h[\mathbf{r}_t\odot\mathbf{h}_{t-1};\mathbf{x}_t]+\mathbf{b}_h\bigr),\\
    \mathbf{h}_t &= (1-\mathbf{z}_t)\odot\mathbf{h}_{t-1} + \mathbf{z}_t\odot\tilde{\mathbf{h}}_t,
\end{align}

where $\mathbf{W}_{(\cdot)}\in\mathbb{R}^{d_h\times(d_h+d_x)}$ and $\mathbf{b}_{(\cdot)}\in\mathbb{R}^{d_h}$ are trainable parameters. The \emph{reset gate} $\mathbf{r}_t$ decides how much past information contributes to the candidate state $\tilde{\mathbf{h}}_t$, while the \emph{update gate} $\mathbf{z}_t$ controls the trade-off between keeping the previous state $\mathbf{h}_{t-1}$ and adopting the new candidate.

For greater representational power, one can stack $L$ identical GRU layers, passing the output of layer~$\ell$ at step~$t$ as input to layer~$\ell\!+\!1$. In multistep forecasting, an \emph{encoder–decoder} architecture is commonly adopted:

\begin{itemize}
  \item \textbf{Encoder} processes a window of $T_e$ historical inputs
        and returns a context vector $\mathbf{h}_{T_e}^{(e)}$.
  \item \textbf{Decoder} starts from this context and autoregressively
        generates a sequence of $T_d$ future outputs $\{\mathbf{y}_1,\dots,\mathbf{y}_{T_d}\}$
\end{itemize}

For practical advantages, compared with more complex gated cells, GRUs require fewer parameters, converge faster, preserve long-term trends better than vanilla RNNs, and meet typical inference budgets for industrial control loops. These properties make GRUs a natural choice for real-time, model-predictive control of thermal-fluid systems.

\begin{table}[htbp]
    \caption{Hyperparameter space for GRU model optimization and training.}
    \label{tab:gru_config}
    \centering
    \small
    \begin{tabular}{@{}>{\raggedright\arraybackslash}p{2.8cm}
                    >{\raggedright\arraybackslash}p{2.8cm}
                    >{\raggedright\arraybackslash}p{2.8cm}
                    >{\raggedright\arraybackslash}p{4.0cm}@{}}
        \toprule
        \textbf{Component} & \textbf{Configuration} & \textbf{Value Space} & \textbf{Rationale} \\
        \midrule
        \multicolumn{4}{l}{\textit{Network Architecture}} \\
        Number of layers & GRU depth & 1, 2, 3 & Determine model capacity \\
        Hidden size & Units per layer & 128, 256, 512, 1024 & Balance representation power and complexity \\
        Input window & Look-back time & 30 & Capture system history (30s) \\
        Output dimension & Prediction time & 10 & Predict the next state and control signal \\
        \midrule
        \multicolumn{4}{l}{\textit{Training Configuration}} \\
        Batch size & Training samples & 128 & Optimize GPU memory \\
        Learning rate & Initial & 1e-3 & Adam optimizer default \\
        Optimizer & Algorithm & Adam & Standard adaptive learning rate \\
        Scheduler & Type & ReduceLROn-Plateau & Adapt learning rate on validation loss plateau \\
        Early stopping & Patience & 100 epochs & Prevent overfitting \\
        \midrule
        \multicolumn{4}{l}{\textit{Loss Function}} \\
        Prediction loss & Metric & MSE & Penalize large errors in state prediction \\
        Regularization & Type / Strength & weight decay = 1e-5 & Mitigate overfitting via weight decay \\
        \bottomrule
    \end{tabular}
\end{table}

Instead of a complex encoder-decoder structure, our study employed a direct sequence-to-sequence GRU architecture, which proved more efficient for this control task. This architecture consists of a multi-layer GRU network followed by a fully-connected linear layer. The GRU network processes a sequence of historical data over a defined input window to capture the system's temporal dynamics. The final hidden state of the GRU is then passed to the linear layer, which maps the learned features to a prediction for the system's state at the next 10-step sequence. To identify the optimal controller, a systematic hyperparameter search was conducted according to the configuration space detailed in Table \ref{tab:gru_config}, exploring architectures with 1 to 3 GRU layers and hidden sizes ranging from 128 to 1024 units.

The training process utilized experimental data collected from various operational scenarios, including steady-state conditions, power ramps, and flow transients. The dataset comprises over 3,706 time steps of synchronized measurements across all sensors. Data preprocessing involved several key steps: 1) Z-score normalization using training set statistics, 2) Sliding window generation with 30-step input (30 sec) and 10-step (10 sec) output sequences, 3) 70/10/20 train/validation/test split with sequential partitioning. Specifically, the model's input consists of a 30-step (30 sec) sequence of 26 variables: 25 historical sensor measurements and a demand for electric power requested by the user or an external energy system. The generated electric power is calculated from the power of the tertiary loop, assuming a typical thermal to electrical conversion efficiency of 45\% in the high-temperature molten salt reactor \citep{zohuri2021molten, forsberg2003molten}. The model is trained to predict a 10-step (10 sec) output sequence of 29 variables: the future values of the 25 system parameters and 4 corresponding control action signals (heater, two pumps, and control rod controllers).

%%%%%%%%%%%%%%%%%%%%%%%%%%%%%%%%%%%%%%%%%%%%%%%%%%%%%%
\subsubsection{prediction of future condition and control actions}
Table~\ref{tab:hp_sweep} summarizes the average training and validation losses obtained for each combination of hidden‑layer width and depth explored in the grid search. As expected, increasing either dimension generally lowers the training loss, yet the validation surface exhibits a shallow U‑shape: two‑layer networks consistently outperform their one‑layer counterparts, whereas a third layer tends to overfit.

\begin{table}[htbp]
    \centering
    \caption{Sensitivity of average training loss and validation loss to GRU depth and hidden size.}
    \label{tab:hp_sweep}
    \small
    \begin{tabular}{ccccc}
        \toprule
        Hidden size & Layers & Train Loss & Valid Loss & Parameters \\
        \midrule
        128  & 1 & 0.1287 & 0.1731 & 0.125 M \\
        \textbf{128}  & \textbf{2} & \textbf{0.1221} & \textbf{0.1661} & \textbf{0.323 M} \\
        128  & 3 & 0.1102 & 0.1757 & 0.521 M\\
        256  & 1 & 0.1250 & 0.1682 & 0.446 M\\
        256  & 2 & 0.1063 & 0.1689 & 1.235 M\\
        256  & 3 & 0.0990 & 0.1698 & 2.025 M\\
        512  & 1 & 0.1181 & 0.1712 & 1.678 M\\
        512  & 2 & 0.1134 & 0.1726 & 4.830 M\\
        512  & 3 & 0.1010 & 0.1708 & 7.982 M\\
        1024 & 1 & 0.1057 & 0.1695 & 6.502 M\\
        1024 & 2 & 0.1015 & 0.1691 & 19.098 M\\
        1024 & 3 & 0.0845 & 0.1701 & 31.693 M\\
        \bottomrule
    \end{tabular}
\end{table}

The numerically best validation score (0.166) is achieved by the 128‑unit, two‑layer network. In practice, however, we adopt the \textbf{256‑unit, two‑layer} configuration for the following reasons: Marginal accuracy trade‑off. The difference in validation loss between 128×2 and 256×2 is only $2.7\times10^{-3}$—well within the uncertainty band introduced by data noise. Greater representational headroom. The wider latent space of the 256‑unit model better captures rare transients (e.g.\ heater saturation) that are under‑represented in the current dataset, Real‑time feasibility. 

\begin{figure}[htbp]
\centering
\includegraphics[width=0.86\textwidth]{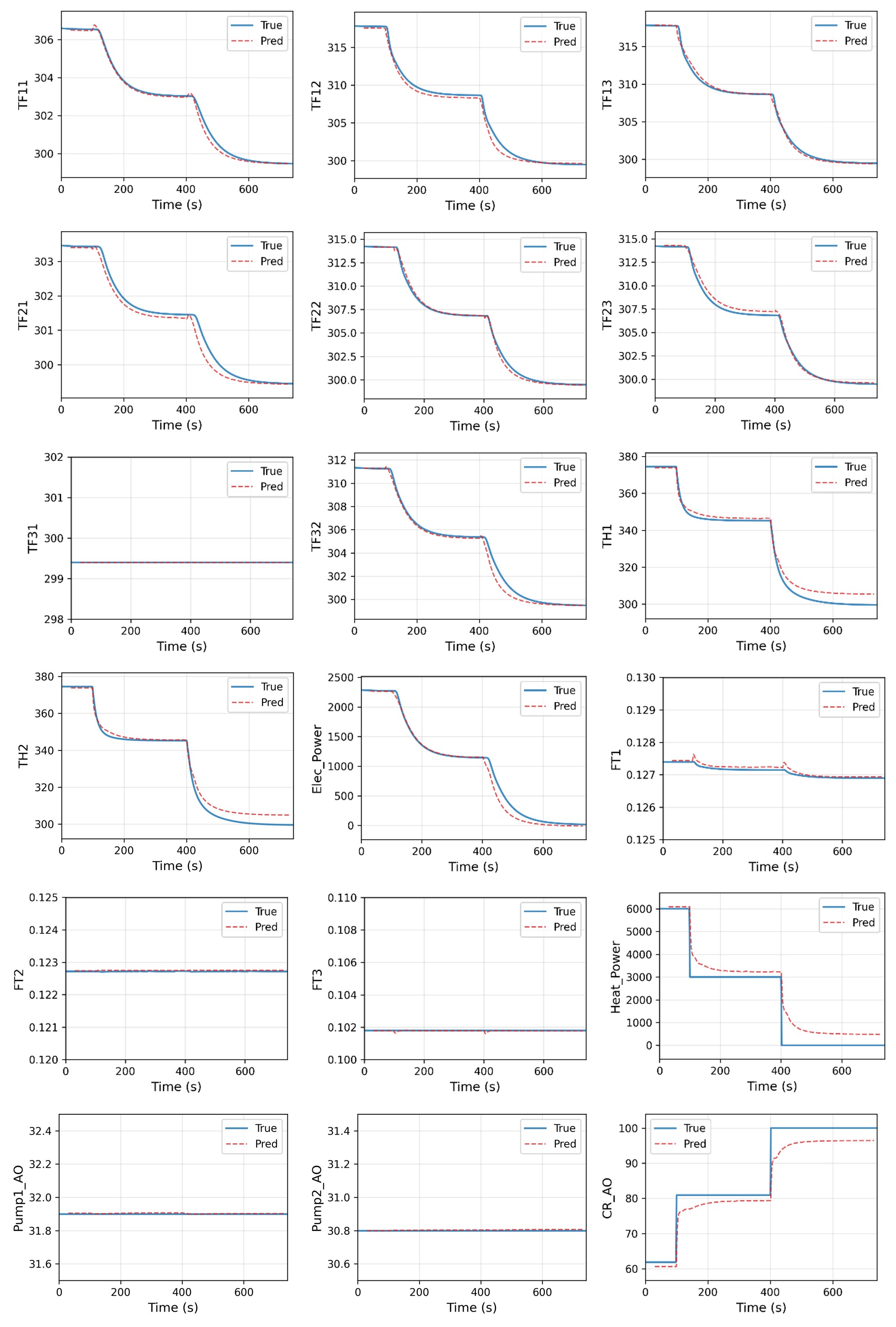}
\caption{Prediction results of optimized GRU model for unseen test dataset: fluid temperatures (TF11, TF12, TF13, TF21, TF22, TF23, TF31, TF32), heater temperature (TH1), heating power (Heat\_Power), estimated electrical power (denoted as Elec\_Power), flow rates (FT1, FT2, FT3), control signals for heater, pump, and control rod position (denoted as Heater\_AO, Pump1\_AO, Pump2\_AO, CR\_AO).}
\label{fig:MLresults}
\end{figure}

Empirically, adding a third layer offered diminishing returns and even degraded validation loss, signalling over‑fitting. Hidden dimensions beyond 512 produced negligible gains at the cost of memory footprint. These findings informed the final deployment choice.

Figure~\ref{fig:MLresults} shows prediction result for the unseen test dataset, superimposing the 10~s–ahead GRU forecasts (red dashed) on the measured responses (blue solid) for all output channels. The transient can be interpreted as occurring in three consecutive phases.

\begin{table}[htbp]
    \centering
    \caption{ML model prediction accuracy over the full test sequence.}
    \label{tab:gru_metrics}
    \small
    \begin{tabular}{lcc}
        \toprule
        Variable group & MAE & RMSE\\ \midrule
        Temperatures (16 channels) & 1.16~K & 1.42~K\\
        Pressures (4 channels) & $1.17\times10^{3}$~Pa & $1.22\times10^{3}$~Pa\\
        Flow rates (3 channels) & $7.7\times10^{-5}$~m$^{3}$\,s$^{-1}$ & $8.3\times10^{-5}$~m$^{3}$\,s$^{-1}$\\
        Power & 0.26~kW & 0.36~kW\\
        Actuators (4 AOs) & 2.76\,\% & 3.82\,\%\\
        \bottomrule
    \end{tabular}
\end{table}

During the \emph{initial power step} (0–200 s) which demand electric power 3.0 kW to 2.27 kW produces an 8–10 K drop in the primary‑side temperatures TF11–TF32 and TH1 (Sensor locations and nomenclature—e.g.\ TF11, TF21—are detailed in Fig.~\ref{fig:sam_nodal}). The model accurately replicates both the steep gradient and the subsequent inflection due to coolant mixing; the maximum instantaneous error remains below 0.5 K.

In the \emph{intermediate ramp} phase (200–420 s, demand electric power: 1.14 kW), a series of smaller heater reductions imposes a staircase profile on 'Heat\_Power' and 'Elec\_Power'. The GRU follows each edge with an average undershoot of about 3\%, a bias attributed to actuator latency effects that are only partially represented in the training data.

During \emph{re‑equilibration} (420–700 s, demand electric power: 0 kW), the thermal variables relax exponentially toward their new steady state, while pressure and flow remain quasi‑steady. Throughout this period, the forecasts stay within the sensors’ noise band, indicating that the learned model preserves the dominant first‑order dynamics even as the prediction horizon is rolled forward repeatedly. The largest discrepancies arise in 'Heater\_AO', whose rare saturation events produce the vertical jumps in row 4 of the figure; these represent fewer than 2\% of training samples and will be up‑weighted in future data‑collection campaigns. Overall, the predictor delivers sub‑percent errors for core safety variables while remaining lightweight enough for real‑time deployment.

Table~\ref{tab:gru_metrics} quantifies the model’s accuracy across the major sensor classes. Temperature channels—the primary safety indicators—are predicted within 1.5 K (RMSE), well around the \SI{1.0} {\celsius} resolution of the Type‑K thermocouples. Pressure and flow‑rate errors correspond to less than 2~\% of their nominal operating ranges, while power estimates remain within 0.4 kW. The slightly higher percentage error for actuator analogue outputs (3–4\%) is dominated by 'Heater\_AO', whose rare saturation events introduce sharp discontinuities that were under-sampled during training. Overall, the close agreement between prediction and measurement across temperature, pressure, flow, and power channels confirms that the GRU captures the essential thermal‑fluid behavior required for real‑time predictive control.

\subsubsection{AI-enabled real-time digital twin operation}

\begin{figure}[htbp]
    \centering
    \includegraphics[width=\textwidth]{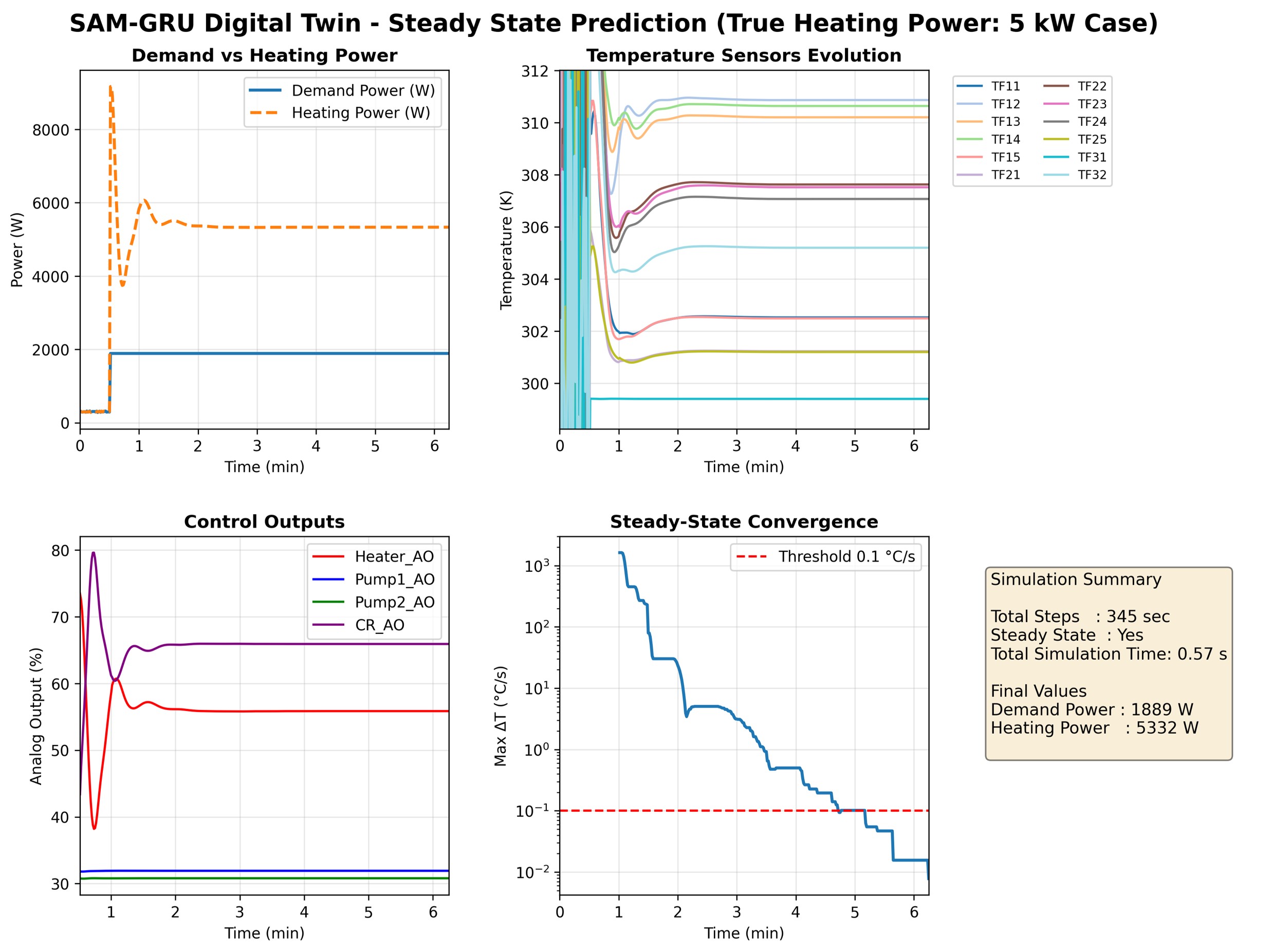}
    \caption{Case study: quasi-real-time AI-driven SAM digital twin's steady-state prediction: predicted heating power of 5,332 W for a 1,889 W demand power (True heating power: 5,000 W, DT duration time: 0.57 sec).}
    \label{fig:DT_steadystate}
\end{figure}

The GRU-accelerated SAM digital twin demonstrates remarkable capability in predicting facility steady-state conditions with high computational efficiency. Figure~\ref{fig:DT_steadystate} presents a comprehensive case study where the digital twin predicts system response to a step change in demand electric power from 0 kW to 1.889 kW, corresponding to a true heating power target of 5 kW.

The digital twin simulation achieved steady-state convergence in just 345 time steps (5.75 minutes of simulated time), executed in merely 0.57 seconds of wall-clock time—representing a speed-up factor of approximately 600× compared to real-time operation. This acceleration enables operators to rapidly evaluate multiple control scenarios and anticipate system behavior well ahead of actual transients.

Analysis of the steady-state predictions reveals generally good agreement with true values, though with systematic biases that merit discussion. The digital twin predicted a heating power of 5,332 W compared to the true value of 5,000 W (6.6\% over-prediction), while the electric power prediction of 1,096 W underestimated the true value of 1,889 W by 42\%. This discrepancy in electric power prediction likely stems from the GRU model's difficulty in capturing the complex nonlinear relationship between heating power and electrical consumption, particularly near operational boundaries.

Temperature predictions demonstrated higher fidelity, with an average absolute error of 3.15 K across all sensors. Individual temperature channels exhibited errors ranging from 0 K (TF31, perfect agreement) to 4.48 K (TF13), all within acceptable bounds considering the Type-K thermocouple uncertainty of ±1.0°C. The model consistently underestimated temperatures by 2–4 K, suggesting a slight conservative bias that enhances operational safety margins. Notably, the heater temperature TH1 was predicted as 362.78 K versus the true value of 366.18 K, maintaining the critical safety-relevant prediction within 3.4 K accuracy.

\begin{figure}[htbp]
    \centering
    \includegraphics[width=\textwidth]{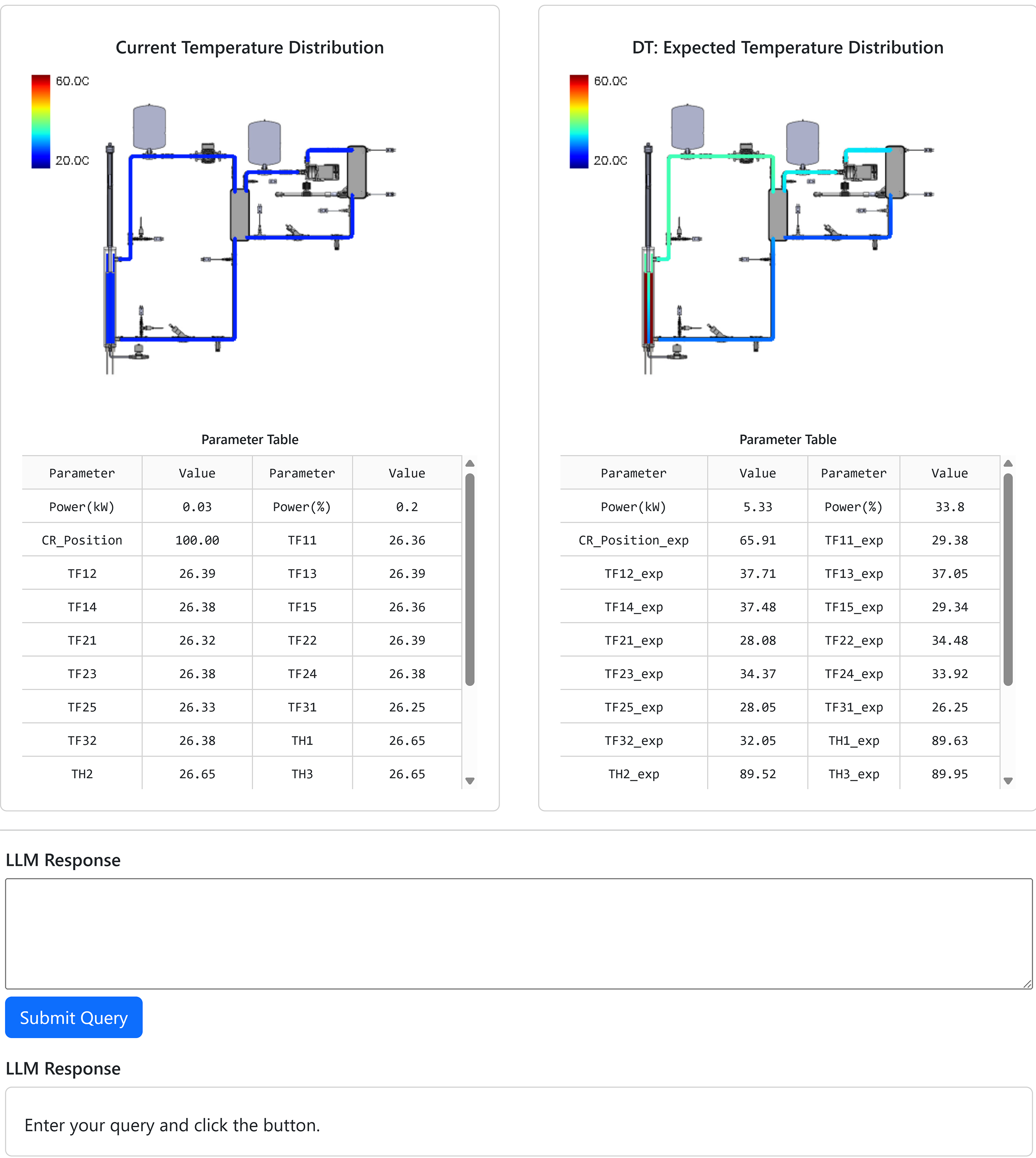}
    \caption{Integrated monitoring dashboard showing real-time facility data readings (left) with predictions of AI-driven SAM digital twin (right) and large language model query (bottom).}
    \label{fig:dashboard}
\end{figure}

Figure~\ref{fig:dashboard} illustrates the integrated monitoring dashboard that leverages these digital twin capabilities for enhanced situational awareness. The left panel displays real-time sensor readings from the physical facility at the current 0 kW operating point, while the right panel visualizes the digital twin's prediction of the temperature distribution following the proposed power increase. This side-by-side comparison enables operators to immediately grasp both current conditions and anticipated future states. The temperature field visualization clearly shows the expected heating patterns, with the primary circuit experiencing temperature rises from the current ~26°C baseline to predicted values ranging from 28–90°C depending on proximity to the heat source.

The dashboard also incorporates a large language model interface (bottom panel) that allows operators to query the system using natural language, further reducing the cognitive burden of interpreting complex thermo-fluid states. This multimodal approach—combining numerical predictions, visual representations, and natural language interaction—exemplifies how ML-enhanced digital twins can transform nuclear facility operations from reactive monitoring to proactive optimization.

The demonstrated prediction accuracy, combined with the 600× computational speedup, validates the GRU-based approach for real-time decision support. While some systematic biases exist (particularly in electric power and heater control predictions), these are well-characterized and can be compensated through operator experience or future model refinements. The ability to evaluate steady-state conditions in under one second opens new possibilities for exploring extensive operational envelopes, optimizing control strategies, and enhancing overall facility safety through predictive analytics.

%%%%%%%%%%%%%%%%%%%%%%%%%%%%%%%%%%%%%%%%%%%%
\subsection{LLM Integration for Operator Assistance}
\label{subsec:llm_case}

To evaluate the effectiveness of LLM integration in testbed operations, we implemented an operator assistance system using OpenAI's GPT-4o~\citep{OpenAI2024GPT4o} model through a LangChain~\citep{H.Chase2023GitHubApplications}-based agent architecture. LangChain facilitates the integration by providing 'ChatPromptTemplate' to structure domain-specific knowledge about our testbed simulator—including operational limits (80°C for coolant loops, 200°C for heater elements) and safety priorities—into a comprehensive system prompt. The 'ChatOpenAI' wrapper connects to the GPT-4o API with a temperature setting of 0.3 for consistent responses.

The implementation utilizes 'AgentExecutor' to orchestrate the processing pipeline: it receives combined inputs of real-time sensor data and digital twin predictions, formats them through the prompt template, and invokes the LLM for analysis. Although the current system operates without external tools (empty tools array), the LangChain architecture provides flexibility for future integration of computational tools or database queries. This modular approach enables seamless updates to the prompt engineering, model selection, or tool integration without restructuring the core application logic.

The LLM receives real-time testbed data through the OPC-UA server, encompassing distinct measurements: temperature sensors (TF11-TF32 for loop temperatures, TH1-TH4 for heater monitoring), pressure transducers (PT1-PT4), flow meters (FT1-FT3), and electrical parameters (voltage, current, power, pump speed, control rod position). The system constructs a comprehensive state representation by combining current measurements with digital twin predictions, enabling comparative analysis between actual and expected system behavior.

The data preprocessing pipeline formats raw sensor values into structured natural language descriptions, maintaining numerical precision while enhancing interpretation ability. For instance, temperature arrays are annotated with their physical locations and operational context, transforming raw values like [26.36, 26.39, 26.39, 26.38, 26.36] into "Primary Loop Temperatures [Testsection Inlet, Testsection Outlet, Top Pump Inlet, 1st-HX Inlet, 1st-HX Outlet]: 26.36°C, 26.39°C, 26.39°C, 26.38°C, 26.36°C."

A comprehensive validation case study was conducted to evaluate the LLM's analytical capabilities and operational guidance quality. The test scenario involved transitioning the system from an idle state to active power generation, representing a common yet critical operational procedure. The validation case study as shown in Table~\ref{testbed-llm-input} and Table~\ref{testbed-llm-output} revealed several key capabilities of the LLM integration:

\textbf{Multi-parameter Correlation Analysis:} The system successfully processed and correlated 31 simultaneous measurements, identifying that zero power output despite electrical readings (35.78 V, 0.72 A) was due to full control rod insertion rather than equipment malfunction. This demonstrates the LLM's ability to apply domain knowledge in interpreting complex system states.

\textbf{Quantitative Accuracy:} The LLM maintained numerical precision throughout its analysis, correctly citing specific values (e.g., "65.91\% control rod position," "37.71°C outlet temperature") and calculating appropriate safety margins. The system demonstrated understanding of operational limits by comparing predicted values against established thresholds.

\textbf{Predictive Reasoning:} By integrating digital twin predictions, the LLM provided forward-looking analysis of system behavior. It correctly identified that an 11°C temperature rise in the primary loop and ~90°C heater temperatures were within acceptable ranges, demonstrating capability for anticipatory safety assessment.

\textbf{Actionable Guidance Generation:} The LLM produced specific, implementable recommendations including stepwise control procedures (5\% increments with 30-second holds), monitoring priorities (TH1-TH4, power-to-flow ratio), and decision points (80\% rod position hold). This level of detail surpasses generic advisory systems.

%========================================
\begin{table}[htbp]
\centering
\caption{LLM validation input of testbed: integration of current measurements and digital twin expectations (Human Message)}
\label{testbed-llm-input}
\begin{scriptsize}
\begin{tabularx}{0.95\textwidth}{|>{\raggedright\arraybackslash}X|}
  \hline
  % Centered header row:
  \multicolumn{1}{|c|}{\textbf{INPUT (Human Message)}}\\
  \hline
  % Body text (left-aligned by raggedright):
  \textbf{Current Facility Data:} \newline
  - Total Heater Voltage: 35.78 V \newline
  - Total Heater Current: 0.72 A \newline
  - Total Power: 0.00 kW (0.0\%) \newline
  - Control Rod Position: 100.00\% \newline
  - Primary Loop Temperatures [Testsection Inlet Temp, Testsection Outlet Temp, Top Pump Inlet Temp, 1st-HX Inlet Temp, 1st-HX Outlet Temp]: 26.36°C, 26.39°C, 26.39°C, 26.38°C, 26.36°C \newline
  - Secondary Loop Temperatures [1st-HX 2nd-side Inlet Temp, 1st-HX 2nd-side Outlet Temp, Top Pump Inlet Temp, 2st-HX Inlet Temp, 2st-HX Outlet Temp]: 26.32°C, 26.39°C, 26.38°C, 26.38°C, 26.33°C \newline
  - Four Heater Temperature in Test Section: 26.65°C, 26.65°C, 26.65°C, 26.65°C \newline
  - Heat Sink Loop Temperatures [HX Inlet Temp, HX Outlet Temp]: 26.25°C, 26.38°C \newline
  - Gauge Pressure [1st Loop Testsection Inlet, 1st Loop Testsection Outlet, 1st Loop Top, 2nd Loop Top]: 112.86 kPa, 106.29 kPa, 100.03 kPa, 100.05 kPa \newline
  - Flow Rate [1st Loop, 2nd Loop, 3rd Loop]: 0.1269 kg/s, 0.1227 kg/s, 0.1018 kg/s \newline
  \newline
  \textbf{Digital Twin's Expectation Data for User's Demand Power: 1.89 kW} \newline
  - Expected Total Power: 5.33 kW (33.8\%) \newline
  - Expected Control Rod Position: 65.91\% \newline
  - Digital Twin Expectation for Primary Loop Temperatures [Testsection Inlet Temp, Testsection Outlet Temp, Top Pump Inlet Temp, 1st-HX Inlet Temp, 1st-HX Outlet Temp]: 29.38°C, 37.71°C, 37.05°C, 37.48°C, 29.34°C \newline
  - Digital Twin Expectation for Secondary Loop Temperatures [1st-HX 2nd-side Inlet Temp, 1st-HX 2nd-side Outlet Temp, Top Pump Inlet Temp, 2st-HX Inlet Temp, 2st-HX Outlet Temp]: 28.08°C, 34.48°C, 34.37°C, 33.92°C, 28.05°C \newline
  - Digital Twin Expectation for Four Heater Temperature in Test Section: 89.63°C, 89.52°C, 89.95°C, 90.19°C \newline
  - Digital Twin Expectation for Heat Sink Loop Temperatures [HX Inlet Temp, HX Outlet Temp]: 26.25°C, 32.05°C \newline
  - Digital Twin Expectation for Gauge Pressure [1st Loop Testsection Inlet, 1st Loop Testsection Outlet, 1st Loop Top, 2nd Loop Top]: 109.67 kPa, 104.36 kPa, 97.67 kPa, 99.56 kPa \newline
  - Digital Twin Expectation for Flow Rate [1st Loop, 2nd Loop, 3rd Loop]: 0.1275 kg/s, 0.1228 kg/s, 0.1018 kg/s \newline
  \newline
  \textbf{User's Question:} Using the current measured data (voltage, current, loop temperatures, pressures, flow rates and control-rod position) and the Digital Twin's predicted behavior (dt\_** values) for a user demand power increase, please 1) Analyze the system's present state and Forecast the system's future state under that user's demand demand increase. 2) Highlight any safety concerns or operational risks that emerge in each case.
  \\
  \hline
\end{tabularx}
\end{scriptsize}
\end{table}

%========================================
\begin{table}[htbp]
    \centering
    \caption{LLM response corresponding to the input query (AI Message)}
    \label{testbed-llm-output}
    \begin{scriptsize}
    \begin{tabularx}{0.95\textwidth}{|X|}
        \hline
        \textbf{Response:} 
        To address the user's question, we need to analyze the current state of the system and forecast its future state under the user's demand for increased power. Here is a step-by-step analysis:\newline
        \newline
        \textbf{Current State Analysis}\newline
        \newline
        1. \textbf{Power and Control Rod Position:}\newline
        - Current Total Power: 0.00 kW (0.0\%) | Current Control Rod Position: 100.00\%\newline
        - Expected Total Power: 5.33 kW (33.8\%) | Expected Control Rod Position: 65.91\%\newline
        The current power output is significantly lower than expected, indicating that the system is not generating the required power. The control rod position is fully inserted, which is not aligned with the expected position for the desired power output.\newline
        2. \textbf{Temperatures:}\newline
        - Primary Loop: Temperatures are significantly lower than expected, indicating insufficient heat generation.\newline
        - Secondary Loop: Similarly, temperatures are lower than expected, suggesting inadequate heat transfer from the primary loop.\newline
        - Heat Sink Loop: Temperatures are close to expected values, indicating that the heat sink loop is functioning normally.\newline
        3. \textbf{Pressures and Flow Rates:}\newline
        - Pressures in the loops are slightly higher than expected, which could be due to the low power output and reduced heat generation.\newline
        - Flow rates are close to expected values, indicating that the pumps are operating correctly.\newline
        4. \textbf{Heater Temperatures:}\newline
        - All four heaters are at 26.65°C, much lower than the expected range of around 89-90°C, indicating that the heaters are not operating at the required power level.\newline
        \newline
        \textbf{Forecast for Increased Power Demand}\newline
        \newline
        1. \textbf{Power Adjustment:}\newline
        - To meet the user's demand for increased power, the control rod position should be adjusted to around 65.91\% to increase the power output to approximately 5.33 kW.\newline
        2. \textbf{Temperature and Pressure Changes:}\newline
        - As power increases, the temperatures in the primary and secondary loops are expected to rise closer to the digital twin's expected values.\newline
        - Pressure may decrease slightly as the system stabilizes with increased power and heat generation.\newline
        3. \textbf{Flow Rates:}\newline
        - Flow rates are expected to remain stable as they are already close to expected values.\newline
        \newline
        \textbf{Safety Concerns and Operational Risks}\newline
        \newline
        1. \textbf{Overheating Risk:}\newline
        - Ensure that the heater temperatures do not exceed their limit of 200°C as power increases.\newline
        - Monitor loop temperatures to ensure they remain below the 80°C limit.\newline
        2. \textbf{Pressure Management:}\newline
        - Ensure that pressures remain within safe operating limits to prevent any risk of leaks or system damage.\newline
        3. \textbf{System Stability:}\newline
        - Gradually adjust the control rod position to avoid sudden changes in power output, which could destabilize the system.\newline
        \newline
        By addressing these concerns and making the necessary adjustments, the system can safely meet the user's increased power demand while maintaining operational stability.
        \\
        \hline
    \end{tabularx}
    \end{scriptsize}
\end{table}

The LLM demonstrated conservative safety bias in its recommendations, consistently prioritizing gradual transitions and comprehensive monitoring over rapid achievement of target conditions. When presented with ambiguous scenarios, the system defaulted to requesting additional information rather than making assumptions, aligning with the nuclear industry's safety culture.

Error handling was validated through the injection of anomalous data patterns. The LLM correctly identified and flagged inconsistent sensor readings (e.g., negative flow rates, temperature inversions) and provided appropriate warnings to operators. This robustness is critical for maintaining trust in automated advisory systems.

This validation demonstrates that LLM integration can enhance operator decision-making without replacing human judgment. The natural language interface reduces cognitive load by presenting complex system states in accessible formats while maintaining technical accuracy. The successful correlation of real-time data with physics-based predictions (digital twin) establishes a new paradigm for intelligent operation support systems.

The case study confirms that properly configured LLMs can serve as effective "digital advisors" in nuclear facilities, providing rapid situation assessment, predictive analysis, and structured operational guidance. This capability is particularly valuable during transient operations, where operators must process multiple data streams while making time-critical decisions.

%%%%%%%%%%%%%%%%%%%%%%%%%%%%%%%%%%%%%%%%%%%%%%%
\subsection{Other Potential Applications}

Beyond direct control and operator assistance, the thermal-fluid testbed is designed to support research and development in several other important areas, including: 

\begin{itemize}
    \item \textbf{Anomaly detection}: By leveraging a digital twin to predict future system states, the framework can identify deviations from expected behavior~\cite{kim2025robust}. This is achieved using advanced machine learning techniques, specifically graph neural networks (GNNs), which infer the overall system status and predict future transients to enable the rapid detection of abnormalities~\cite{liu2024development}.
    \item \textbf{Thermal-fluid}: The testbed also serves as a valuable platform for cybersecurity research. We can use GNNs to detect cyber threats such as false data injection attacks by analyzing the learned relationships between different sensor readings~\cite{liu2024development, yoo2025self}. To aid human operators in responding to such threats, an explainer module is being developed to interpret the GNN's output and clarify the nature of the detected cyber attack.
    \item \textbf{Predictive maintenance}: The framework can be extended for predictive maintenance. Through the continuous monitoring and analysis of system data, it becomes possible to forecast potential equipment failures before they occur~\cite{zhao2021prognostics}. This predictive capability allows maintenance to be scheduled proactively, which is crucial for improving the long-term safety, reliability, and operational availability of the reactor system.
\end{itemize}
%%%%%%%%%%%%%%%%%%%%%%%%%%%%%%%%%%%%%%%%%%%

\section{Conclusions}
\label{sec:conclusions}
In this work, we developed and validated a multipurpose, artificial intelligence (AI)-driven thermal-fluid testbed, establishing a new integrated paradigm for advancing small modular reactor technologies. The principal achievement is the seamless fusion of a physical experimental facility, a high-fidelity digital twin, and advanced AI technologies into a single, cohesive framework. This platform closes a gap between theoretical modeling and practical implementation, creating a versatile environment for comprehensive nuclear thermal-fluid research, including digital twinning, autonomous control, and real-time operator assistance. The robust integration, enabled by industrial-grade communication protocols, proves that complex physical and digital systems can operate and collaborate in real-time.

The key scientific findings of this study demonstrate the transformative capabilities of this integrated approach. First, the physics-based digital twin, validated against experimental data, exhibited exceptional accuracy, predicting system temperatures with errors below one degree Kelvin. This high-fidelity model serves as a reliable virtual representation of the facility. Second, a Gated Recurrent Unit (GRU) neural network model was successfully deployed for two critical functions: it accurately forecasted future system states and control actions with a temperature prediction error of approximately 1.4 K, and it accelerated the digital twin’s simulations by a factor of 600, enabling faster than real-time dynamic prediction and analysis. Finally, the integration of a large language model (LLM) created an intelligent operator assistant capable of synthesizing complex data from both the physical facility and the digital twin, delivering actionable insights and safety recommendations in natural language.

Beyond its immediate applications in control and operator assistance, the testbed is designed to support several other critical research areas. Its predictive capabilities are ideal for anomaly detection, as the digital twin's forecast of expected behavior can be compared against real-time data to instantly flag deviations. This same analytical framework makes the platform a valuable tool for cybersecurity research, using machine learning models like graph neural networks to identify sophisticated threats such as false data injection attacks. Furthermore, the continuous monitoring of integrated system data enables powerful predictive maintenance strategies, allowing for the proactive forecasting of potential equipment failures to improve long-term safety, reliability, and operational availability.

The demonstrated platform provides a tangible pathway toward the realization of autonomous control and intelligent operation for advanced reactors, which is essential for enhancing safety, flexibility, and economic competitiveness. By creating a holistic environment where AI-driven control strategies can be developed, tested, and validated against a physical system, this work accelerates the research and development lifecycle. This AI-enhanced, cyber-physical infrastructure represents a foundational tool that will help propel the next generation of smart, reliable, and efficient nuclear systems from concept to deployment.

\section*{Nomenclature}
\begin{tabular}{@{}ll}
    \textbf{Symbol} & \textbf{Description (units)}\\
    $T$                 & Temperature (\si{\celsius})\\
    $Q$                 & Heater power (kW)\\
    $\dot{m}$           & Mass-flow rate (\si{\kilo\gram\per\second})\\
    $P$                 & Pressure (\si{\kilo\pascal})\\
    $\sigma(\cdot)$     & Logistic‐sigmoid activation function\\
    $\mathbf{x}_t$      & Input feature vector at time step $t$\\
    $\mathbf{h}_t$      & GRU hidden state at time step $t$\\
    $\mathbf{r}_t$      & GRU reset gate \\
    $\mathbf{z}_t$      & GRU update gate \\
    $\tilde{\mathbf{h}}_t$ & GRU candidate hidden state \\
    $L$                 & Number of stacked GRU layers\\
    $d_x$               & Dimension of the input vector\\
    $d_h$               & Dimension of the hidden state (``hidden size'')\\
    $T_e$               & Encoder look-back window (time steps)\\
    $T_d$               & Decoder prediction horizon (time steps)\\
\end{tabular}

\section*{Abbreviations}
\begin{tabular}{@{}ll}
AI & Artificial Intelligence\\
CA & Control Algorithm\\
DT & Digital Twin\\
GRU   & Gated Recurrent Unit\\
LLM & Large Language Model\\
ML & Machine Learning\\
MPC   & Model-Predictive Control\\
OPC-UA& Open Platform Communications – Unified Architecture\\
PID   & Proportional–Integral–Derivative (controller)\\
P\&ID & Process and Instrumentation Diagram\\
PLC & Programmable Logic Controller\\
SAM & System Analysis Module\\
SMR & Small Modular Reactor\\
\end{tabular}

\section*{Acknowledgements}
This work is supported by the US Department of Energy Office of Nuclear Energy Distinguished Early Career Program under contract number DE-NE0009468.

\section*{Data Availability}
The data presented in this study are available from the corresponding author upon reasonable request. All requests are subject to review prior to data release.

\section*{Conflicts of Interest}
The authors declare no conflicts of interest.

%% bibliography style file
\bibliographystyle{elsarticle-num}

\bibliography{references}

\end{document}